\documentclass[prc,twocolumn,superscriptaddress,nofootinbib]{revtex4-1}
\usepackage{newtxtext}
\usepackage[varvw,bigdelims]{newtxmath}
\usepackage[colorlinks=true,allcolors=blue]{hyperref}
\usepackage{graphicx}
\usepackage{bm}
\usepackage{amsmath}
\usepackage{xcolor}

\begin{document}
\title{\boldmath 
  Correlation function for the $n \bar D_{s0}^*(2317)$ interaction and the issue of elastic unitarity}

\author{Natsumi Ikeno}
\email{ikeno@maritime.kobe-u.ac.jp}
\affiliation{Graduate School of Maritime Sciences, Kobe University, Kobe 658-0022, Japan}

\author{Eulogio Oset}
\email{Eulogio.Oset@ific.uv.es}
\affiliation{Departamento de F\'{i}sica Teórica and IFIC, Centro Mixto Universidad de Valencia-CSIC, Institutos de Investigaci\'{o}n de Paterna, Aptdo. 22085, E-46071 Valencia, Spain}
\affiliation{Guangxi Key Laboratory of Nuclear Physics and Technology, Guangxi Normal University, Guilin 541004, China}%

\begin{abstract}
 We study the interaction of a neutron with the $\bar D_{s0}^*(2317)$ resonance and look at the amplitude below threshold and close above threshold. The study is done from the perspective that the $D_{s0}^*(2317)$  resonance is a molecular state of $DK$ in $I=0$. To study this interaction, we use the Fixed Center Approximation to Faddeev equations that considers the $DK$ molecule as the cluster and the neutron as the external particle. We improve the Fixed Center approach to implement elastic unitarity around threshold, which is needed to obtain scattering parameters and to evaluate the $n \bar D_{s0}^*(2317)$ correlation function that we determine here. One interesting result of the study is the appearance of a resonant state below threshold with a binding of about 130 MeV and a width of about 80 MeV, which we suggest to look at in reactions measuring the invariant mass of $\pi \Sigma \bar D$. The ALICE collaboration has initiated studies of this type, by looking at the $p f_1(1285)$ correlation function, and we can only encourage work in this direction which should provide much valuable information on the nature of many resonant states.  
\end{abstract}

\date{\today}

% \pacs{21.65.Ef, 25.70.-z, 25.80.Ls}
\maketitle
%------------------------------------------------------------
\section{Introduction}
Correlation functions are gradually catching up as a tool to learn about hadron interactions, particularly for pairs of hadrons which cannot be formed in scattering experiments~\cite{Fabbietti:2020bfg,ALICE:2020mfd}. 
Theoretical work on this issue follows~\cite{Morita:2014kza,Ohnishi:2016elb,Sarti:2023wlg,Morita:2016auo,Hatsuda:2017uxk,Mihaylov:2018rva,Haidenbauer:2018jvl,Morita:2019rph,Kamiya:2019uiw,Kamiya:2021hdb,Kamiya:2022thy,Liu:2023uly,Vidana:2023olz,Albaladejo:2023pzq,Liu:2023wfo,Liu:2022nec,Torres-Rincon:2023qll,Ikeno:2023ojl,Molina:2023jov,Molina:2023oeu,Liu:2024uxn,Encarnacion:2024jge,Albaladejo:2023wmv,Ikeno:2025kwe}.
The combination of data from correlation functions and spectroscopy measurements has proved particularly useful to learn about hadron dynamics~\cite{Feijoo:2024qqg}. A recent update of theoretical and experimental papers on the issue can be seen in Refs.~\cite{Encarnacion:2024jge,Encarnacion:2025lyf}. Given the success of two body correlations, the idea has been extended to the study of three body correlations~\cite{DelGrande:2021mju,ALICE:2022boj,ALICE:2023gxp,ALICE:2023bny,Garrido:2024pwi,Garrido:2025lar}. While certainly some dynamical information can be obtained from these studies, we take advantage to warn about the goal of finding the three body forces from there, since they are not an observable, in the sense that they are tied to the implicit off shell dependence of the two body amplitudes of the model used to analyze the data. This has been discussed in Ref.~\cite{Encarnacion:2025lyf} and the extreme model dependence of the induced three body force has been shown in Ref.~\cite{Epelbaum:2025aan} (see also Refs.~\cite{Khemchandani:2008rk,MartinezTorres:2008gy}). Yet, we find quite instructive the study of the correlation function of a particle, and a resonance, particularly if this resonance is a dynamically generated state from the interaction of two other particles. Work in this direction has already begun with the measurement of the correlation function of a proton and the $f_1(1285)$~\cite{Otoninfo}.   The $f_1(1285)$ state appears as a dynamically generated state from the interaction of $K^*\bar{K}-\bar{K}^*K$ in isospin $I = 0$~\cite{Lutz:2003fm,Roca:2005nm,Garcia-Recio:2010enl,Zhou:2014ila,Geng:2015yta,Lu:2016nlp}, and experimentally it is observed from a cluster of $K \bar K \pi$~\cite{Otoninfo}.

  In Ref.~\cite{Encarnacion:2025lyf} the theoretical study of the correlation function for the $p f_1(1285)$ system was undertaken, and it was found to have some sizeable structure, starting from a value around 0.4 at threshold and converging smoothly to 1 around 300~MeV/$c$. The system was found to develop a bound state around 40~MeV below threshold, with a width of about 70~MeV, stemming mostly from the decay of $p \bar K$ to $\pi \Sigma$. Most notably, it was also shown that the correlation function and the possible state below threshold were quite different if the $f_1(1285)$ resonance were an ordinary meson state rather than a molecule of  $K^*\bar{K}-\bar{K}^*K$ nature. 
In Ref.~\cite{Encarnacion:2025lyf}, the correlation function was obtained using the scattering matrices of a proton with a cluster of $K^* \bar K$, $\bar K^* K$ nature, which were obtained using the fixed center approximation (FCA) to the Faddeev equations~\cite{Foldy:1945zz,Brueckner:1953zz,Brueckner:1953zza,Chand:1962ec,Barrett:1999cw,Deloff:1999gc,Kamalov:2000iy}. One of the issues raised in Ref.~\cite{Encarnacion:2025lyf} was the one of the elastic unitarity of the three body amplitude obtained. It was shown to be satisfied only approximately, but it was restored by renormalizing the amplitude with a factor close to unity. In the present work we shall devote some attention to this problem and provide a general framework which respects the elastic unitarity, and apply the method to a different problem, the interaction of a neutron with the $\bar D_{s0}^*(2317)$ state. 
 
 The $D_{s0}^*(2317)$ state is again an example of a dynamically generated state. It is generated from the channels $D^0 K^+$, $D^+ K^0$, and $D^+_{s} \eta$ \cite{vanBeveren:2003kd,Barnes:2003dj,Chen:2004dy,Kolomeitsev:2003ac,Gamermann:2006nm,Guo:2006rp,Yang:2021tvc,Liu:2022dmm}, being mostly a $D K$ state in isospin $I=0$. This is also supported by lattice
 QCD simulations~\cite{Mohler:2013rwa,Lang:2014yfa,Bali:2017pdv,Cheung:2020mql}, and a detailed analysis of lattice QCD  data done in Ref.~\cite{MartinezTorres:2014kpc} quantifies the probability of $DK$ in the wave function of the $D_{s0}^*(2317)$ around 72\%. The state is bound by about 42~MeV with respect to the $D^0 K^+$ threshold and decays to the $D^+_s \pi^0$ channel~\cite{BaBar:2003oey}, an isospin violating mode, as a consequence of which the width is extremely small ($< 3.8$~MeV).

  We have chosen the $n \bar D_{s0}^*(2317)$ instead of the $p D_{s0}^*(2317)$, for two reasons. One is technical, since the system chosen has no Coulomb interaction, which simplifies the formalism, and more important, we have the $n \bar{K}$ interaction which is attractive and leads to the $\Lambda(1405)$. Then we can hope to get some bound state of $n \bar D_{s0}^*(2317)$ which will have repercussions on the correlation function of this system, reflecting the molecular nature of the  $D_{s0}^*(2317)$ state. 
We should note that the system $n \bar D \bar K$ has been studied in Ref.~\cite{Yamagata-Sekihara:2018gah} and a bound state of this system is found. Our aim is different since we want to evaluate the correlation function of $n \bar{D}^*_{s0}(2317)$ and the experimentalist isolates the $\bar D^*_{s0}(2317)$ and the neutron, so we are dealing with a particular configuration in the whole $n \bar D \bar K$ system and study it above the $n \bar{D}^*_{s0}(2317)$ threshold.

This paper is organized as follows. In Sec.~\ref{sec:formalism}, we explain the improved formalism of the Fixed Center Approximation (FCA) to implement elastic unitarity around threshold and to evaluate the $n \bar D_{s0}^*(2317)$ correlation function. In Sec.~\ref{sec:result}, we show our numerical results, and Sec.~\ref{sec:discussion} is devoted to conclusions.

\section{Formalism}\label{sec:formalism}

\subsection{Fixed Center Approximation (FCA)}
The FCA has been often used to study three body systems and in some cases it has been tested with results using Faddeev or variational calculations aiming at getting bound states of the systems. For instance, the $DND^*$ state has been studied in Ref.~\cite{Luo:2022cun} using the Gaussian expansion method~\cite{Hiyama:2003cu,Hiyama:2012sma}, and in Ref.~\cite{Montesinos:2024eoy} using the fixed center approximation (FCA), with similar results. Similarly, the $D^*D^*D^*$ system has been studied in Ref.~\cite{Luo:2021ggs} using the Gaussian expansion method and in Ref.~\cite{Bayar:2022bnc} with the FCA with similar results, and also similar to those found in Ref.~\cite{Ortega:2024ecy} using the ladder amplitude formalism. The limits to the use of FCA can also be seen when applied to obtain states above threshold~\cite{MartinezTorres:2010ax}. More information on the issue can be found in the review paper~\cite{MartinezTorres:2020hus}.

\begin{figure*}[!htb]
\begin{center}
\includegraphics[width=0.7\linewidth]{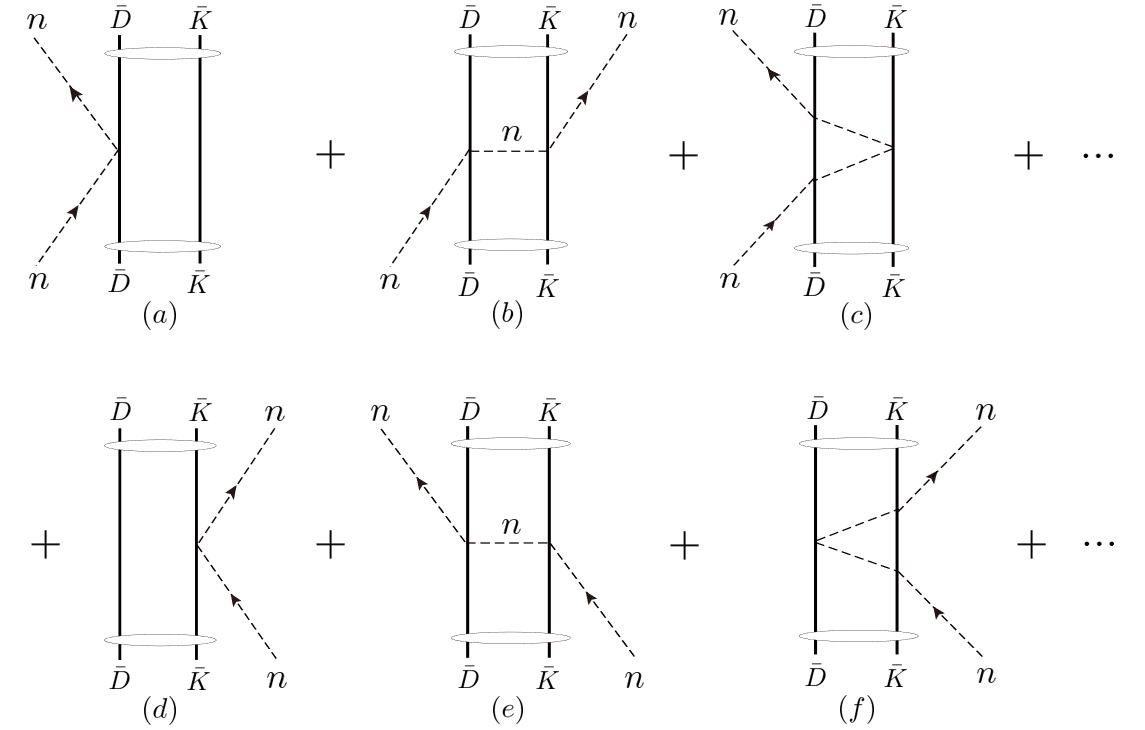}
\caption{Diagrams appearing in the FCA to the $n(\bar{D}\bar{K})_{\text{cluster}}$ interaction.}
\label{fig:1}
\end{center}
\end{figure*} 

In the FCA to $n (\bar{D} \bar{K})_{\text{cluster}}$, with the cluster in $I=0$, we have the diagrams as shown in Fig.~\ref{fig:1}.
Taking into account that the $\bar{D}\bar{K}$ has isospin $I=0$, the $n\bar{D}$ interaction in Fig.~\ref{fig:1}~(a) is given in Refs.~\cite{Roca:2010tf,Xiao:2011rc,Ikeno:2022jbb,Bayar:2023itf,Encarnacion:2025lyf} as
\begin{equation}
\begin{aligned} 
t_{1} &= \frac{3}{4}\,t_{n\bar{D}}^{(1)} + \frac{1}{4}\,t_{n\bar{D}}^{(0)}, \\
t_{2} &= \frac{3}{4}\,t_{n\bar{K}}^{(1)} + \frac{1}{4}\,t_{n\bar{K}}^{(0)},
\end{aligned}
\label{eq:t1t2}
\end{equation}
where $t_{n\bar{D}}^{(I)}$ is the scattering matrix of $n \bar D$ in isospin $I$, and
$t_{n \bar{K}}^{(I)}$ the $n \bar K$ scattering matrix in the isospin $I$.
We show these matrices in Appendix~\ref{sect:app}.

We divert a bit from the formalism of Ref.~\cite{Encarnacion:2025lyf} and write four partition functions, $T_{ij}$ with $i,j = 1,2$.
$T_{11}$ sums all the diagrams where the neutron collides first with particle~1 of the cluster (the $\bar{D}$) and finishes with a collision with particle~1.
$T_{12}$ sums all diagrams where the first collision is with particle~1 and the last one with particle~2 (the $\bar{K}$).
$T_{21}$ sums all diagrams where the first collision is with particle~2 and the last one with particle~1 and, finally, in $T_{22}$ the first collision is with particle~2 and the last one with particle~2.
A similar splitting of the partition functions was done in Ref.~\cite{Sekihara:2016vyd}.

With the previous definitions, it is easy to see the coupled equations:
\begin{equation}
\begin{aligned}
T_{11} &= t_1 + t_1\,\tilde{G}\,T_{21},\\
T_{12} &= t_1\,\tilde{G}\,T_{22},\\
T_{21} &= t_2\,\tilde{G}\,T_{11},\\
T_{22} &= t_2 + t_2\,\tilde{G}\,T_{12},
\end{aligned}
\label{eq:coupled}
\end{equation}
where $\tilde{G}$ in the nucleon propagator folded with the cluster wave function that we detail below.
Eqs.~\eqref{eq:coupled} have a trivial solution as
\begin{equation}
\begin{aligned}
&  T_{11}= \frac{t_1}{1 - t_1 t_2 \,\tilde{G}^{\,2}},
  \qquad
  T_{22}= \frac{t_2}{1 - t_1 t_2 \,\tilde{G}^{\,2}}, \\
&  T_{12}=T_{21}= 
  \frac{t_1 t_2 \,\tilde{G}}{1 - t_1 t_2 \,\tilde{G}^{\,2}}.
\end{aligned}
\label{eq:t11t22t12}
\end{equation}
The total scattering amplitude is given by
\begin{equation}
  T = T_{11}+T_{22}+T_{21}+T_{12} =
  \frac{t_1 + t_2 + 2\,t_1 t_2 \,\tilde{G}}%
       {1 - t_1 t_2 \,\tilde{G}^{\,2}} .
\label{eq:total_T}
\end{equation}
This is the same expression obtained in Refs.~\cite{Roca:2010tf,Xiao:2011rc,Encarnacion:2025lyf},
but the new separation shows the meaning of each term in the total $T$ matrix of Eq.~\eqref{eq:total_T}.

Once again, following the formalism of Refs.~\cite{Roca:2010tf,Xiao:2011rc,Encarnacion:2025lyf},
we implement weight factors to go from the $n\bar{D}$, $n\bar{K}$ frames where the $S$ matrix is written in terms of $t_1$, $t_2$ amplitudes, to the frame where the $S$ matrix is written in terms of the neutron and the cluster.
This is easily implemented by substituting
\begin{equation}
  t_1 \rightarrow
  \tilde{t}_1 = \frac{M_c}{M_{\bar{D}}}\,t_1,
  \quad
  t_2 \rightarrow
  \tilde{t}_2 = \frac{M_c}{M_{\bar{K}}}\,t_2,
\label{eq:t1t2_frame}
\end{equation}
and
\begin{equation}
  \tilde{G}(\sqrt{s}) =
  \int \frac{d^3\bm{q}}{(2\pi)^3}
         \frac{M_N}{E_N(\bm{q})}
         \frac{1}{2\omega_c(\bm{q})}
         \frac{F_c(q)}{\sqrt{s} - E_N(\bm{q}) - \omega_c(\bm{q}) + i\epsilon}
         %F_c(q).
\label{eq:G_tilde}
\end{equation}
where $M_{N}$ and $M_{c}$ are the masses of the neutron and the cluster, respectively,
$E_{N}(\bm{q})=\sqrt{M_{N}^{2}+\bm{q}^{2}}$, $\omega_{c}(\bm{q})=\sqrt{M_{c}^{2}+\bm{q}^{2}}$.
In Eq.~\eqref{eq:G_tilde}, $F_{c}(q)$ is the form factor of the cluster normalized as $F_{c}(0)=1$, given by
\begin{equation}
  F_{c}(q)=\frac{F(q)}{N},
\label{eq:form_factor0}
\end{equation}
with 
\begin{equation}
\begin{aligned}
F(q) = 
\int_{|\bm{p}|<q_{\max},\; |\bm{p}-\bm{q}|<q_{\max}}  & \frac{d^{3}\bm{p}}{(2\pi)^{3}}\;
      \frac{1}
           {M_{c}-\omega_{\bar{D}}(\bm{p})-\omega_{\bar{K}}(\bm{p})} \\
&\cdot      \frac{1}
           {M_{c}-\omega_{\bar{D}}(\bm{p}-\bm{q})-\omega_{\bar{K}}(\bm{p}-\bm{q})},
\end{aligned}
\label{eq:form_factor}
\end{equation}
and
\begin{equation}
  N = F(0) = \int_{|\bm{p}|<q_{\max}} \frac{d^3 \bm{p}}{(2\pi)^3} \left( \frac{1}{M_c-\omega_{\bar{D}}(\bm{p})-\omega_{\bar{K}}(\bm{p})} \right)^2 ,
\end{equation}
where $q_{\max}$ is the regulator in the loop functions used in the dynamical generation of the $D_{s0}^{*}(2317)$ from $DK$. In Ref.~\cite{Ikeno:2023ojl}, we found $q_{\max}= 710$~MeV that produced the right binding of the state.
There is also a small change in Eq.~\eqref{eq:G_tilde} with respect to former works, where a global factor $(2M_{c})^{-1}$ is substituted by $(2\omega_{c}(q))^{-1}$ inside the integral and also $M_c \to \omega_{c}(q)$ in the neutron propagator to take into account recoil corrections, normally  ignored in the FCA.
The $T$-matrix is then written as
\begin{equation}
  \tilde{T} =\frac{\tilde{t}_{1}+ \tilde{t}_{2}+2\, \tilde{t}_{1} \tilde{t}_{2}\,\tilde{G}}
         {1-\tilde{t}_{1} \tilde{t}_{2}\,\tilde{G}^{\,2}}. 
\end{equation}
We also need the arguments of the $t_{1}$ and $t_{2}$ matrices.
Following Refs.~\cite{Roca:2010tf,Encarnacion:2025lyf} we write
\begin{equation}
\begin{aligned}
  s_{1}(n\bar{D})
      &=M_{N}^{2}+(\xi\,m_{\bar{D}})^{2}
        +2\,\xi\,m_{\bar{D}}\,q^{0},\\
  s_{2}(n\bar{K})
      &=M_{N}^{2}+(\xi\,m_{\bar{K}})^{2}
        +2\,\xi\,m_{\bar{K}}\,q^{0},
\label{eq:s1s2}
\end{aligned}
\end{equation}
with $q^{0}$ the neutron energy in the cluster rest frame,
\begin{equation}
  q^{0}
   =\frac{s-M_{n}^{2}-M_{c}^{2}}{2\,M_{c}},\qquad
  \xi=\frac{M_{c}}{M_{\bar{D}}+m_{\bar{K}}},
\end{equation}
where the factor $\xi$ is introduced to account for the binding of the $\bar D \bar K$ state, assuming the binding in $\bar{D}$ and $\bar K$ proportional to their masses.
We can see that $\xi\,m_{\bar{K}} + \xi\,m_{\bar{D}} = M_c$.
The variables $s_1$, $s_2$ of Eq.~\eqref{eq:s1s2} correspond to the invariant mass squared of the $n \bar{D}$ and $n \bar{K}$ systems, respectively.
They are derived from, for instance, $s_2 = (p_N + p_{\bar{K}})^2 = M^2_N + (\xi\,m_{\bar{K}})^2 + 2\,\xi\,m_{\bar{K}}\,q^{0}$.
We have assumed the mass of the $\bar D$ or $\bar K$ reduced by the binding. The evaluation of the invariant $s_1$, $s_2$ is done in the cluster rest frame, for convenience, where the products $\bm{p}_N \cdot \bm{p}_{\bar{D}}$ and $\bm{p}_N \cdot \bm{p}_{\bar{K}}$ vanish for the $s$-wave function of the cluster.

Note that $\tilde{G}(\sqrt{s})$ in Eq.~\eqref{eq:G_tilde} is just the ordinary $G$ function for the propagation of a neutron and a cluster, which is regularized by the form factor $F_{c}(\bm{q})$, which provides a natural convergence factor for the loop function.

\subsection{Elastic unitarity of the $\tilde{T}$ matrix}
Take an extreme case where $\tilde{t}_2$ is large and $\tilde{t}_1$ negligible.  
We would have
\begin{equation}
  \tilde{T} = \tilde{t}_2.
\label{eq:T_extreme}
\end{equation}
While this might be a good approximation for $\tilde{T}$ at the $n \bar{D}_{s0}^{*}(2317)$ threshold and below, let us see what happens at threshold.  
With the Mandl and Shaw normalization for meson and baryon fields that we follow~\cite{MandlShaw}, we have
\begin{equation}
  \tilde{T} =  -\,\frac{8\pi\sqrt{s}}{2M_{N}}\;f^{\text{QM}},
\end{equation}
where $f^{\text{QM}}$ is the standard Quantum Mechanics form of the scattering amplitude.  
Then
\begin{equation}
  - \frac{8\pi\sqrt{s}}{2M_{N}}\, \tilde{T}^{-1} \equiv (f^{\text{QM}} )^{-1}
  \simeq  -\frac{1}{a} +\frac{1}{2}\,r_{0}\,q_{\text{cm}}^2
  -i\,q_{\text{cm}},
\label{eq:scat_eff}
\end{equation}
with $q_{\mathrm{cm}}$ the neutron momentum in the $n \bar{D}_{s0}^{*}$ rest frame.  
Eq.~\eqref{eq:scat_eff} allows us to obtain the scattering length and effective range for the $n\bar{D}_{s0}^{*}$ collision.
The elastic unitarity resides in the factor $-i q_{\text{cm}}$ of Eq.~\eqref{eq:scat_eff}.
 
Eq.~\eqref{eq:scat_eff} using the extreme case of Eq.~\eqref{eq:T_extreme} can give a reasonable value for $a$, but it does not provide the factor $-i q_{\text{cm}}$, and then we cannot rely either on the term $q_{\text{cm}}^{2}$ that provides the effective range. Then, we can also not trust the correlation functions that we would obtain from this $T$-matrix.
The rescue to the problem comes from nuclear physics.  In the interaction of a particle with nuclei, ignoring the interaction of the external particle with two nucleons (this would correspond to ignoring, in our approach, the $\tilde{t}_{1}\tilde{t}_{2}\tilde{G}$ term) one obtains the particle nucleus ``optical potential’’ as $t\rho$, with $t$ the scattering matrix of the external particle with the nucleons and $\rho$ the nuclear density. This, up to a normalization, can be interpreted as
$t_{1}+t_{2}+\cdots + t_N $.  
In our case of two particles, this would be $\tilde{t}_{1}+\tilde{t}_{2}$, which is what we get for $\tilde{T}$ in this case. But once the optical potential is obtained, one uses it within the Schr\"{o}dinger equation to obtain the particle-nucleus scattering matrix~\cite{Ericson:1988gk,Seki:1983sh,Nieves:1993ev,Brown:1975di}.  
This means that in our approach we still have to consider the coherent  $n\bar{D}_{s0}^{*}$  propagation.
This is done in the following way, accounting also for the multiple rescattering of the particles.  
We should note that the terms $\tilde{t}_{1}\tilde{t}_{2}\tilde{G}$ already contain some $n\bar{D}_{s0}^{*}$ propagation through the $\tilde{G}$ function.
This made the $\tilde{T}$ matrix in Ref.~\cite{Encarnacion:2025lyf} nearly unitary, and exact unitarity was fulfilled with a renormalization factor close to~1.
Here, we provide a more accurate and general approach.

\begin{figure*}[!ht]
\begin{center}
\includegraphics[width=0.78\linewidth]{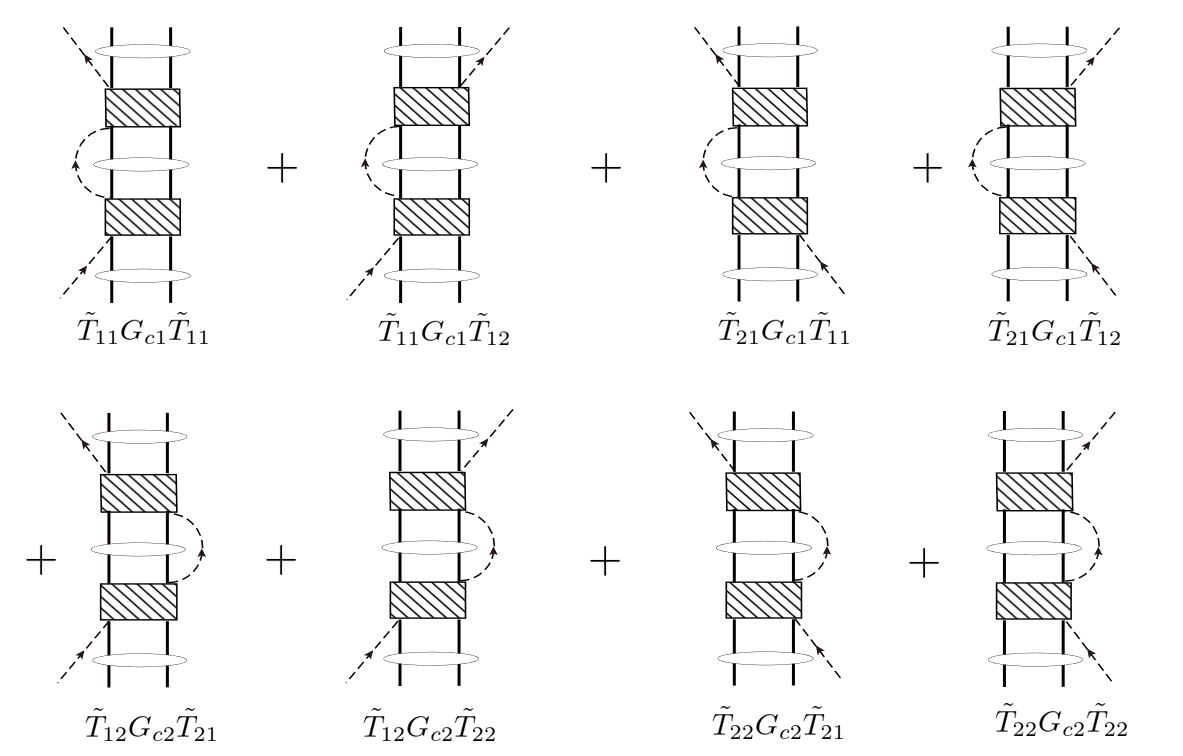} %{fig2_oset.JPG}
\caption{
Terms entering the coherent propagator of the $n\bar{D}_{s0}^{*}$.}
\label{fig:2}
\end{center}
\end{figure*}

To consider the coherent propagation of the $n\bar{D}_{s0}^{*}$ system, we would have the diagrams shown in Fig.~\ref{fig:2}.
We define $\tilde{T}_{ij}$ as in Eq~\eqref{eq:t11t22t12} with $t_i \to \tilde{t}_i$.
In Fig.~\ref{fig:2} we implement the coherent propagation of the $\tilde{T}_{ij}$ terms obtained before.  
We omit the terms $\tilde{T}_{ij} G_{c} \tilde{T}_{k\ell}$ with $j\neq k$ because this involves the neutron going from particle~1 to~2 (or 2 to 1) in the cluster, which is already accounted for by the $\tilde{t}_{1}\tilde{t}_{2} \tilde{G}$ terms of $\tilde{T}$.
The sum of the terms of Fig.~\ref{fig:2} is given by
\begin{equation}
\begin{aligned}
  \tilde{T}_{\text{sum}} =
    & \, \tilde{T}_{11}G_{c1}\tilde{T}_{11}
    +\tilde{T}_{11}G_{c1}\tilde{T}_{12}
    +\tilde{T}_{21}G_{c1}\tilde{T}_{11}
    +\tilde{T}_{21}G_{c1}\tilde{T}_{12} \\
    &+\tilde{T}_{12}G_{c2}\tilde{T}_{21}
    +\tilde{T}_{12}G_{c2}\tilde{T}_{22}
    +\tilde{T}_{22}G_{c2}\tilde{T}_{21}
    +\tilde{T}_{22}G_{c2}\tilde{T}_{22},
\end{aligned}
\label{eq:T_sum}
\end{equation}
where $G_{c1}(\sqrt{s})$ and $G_{c2}(\sqrt{s})$ are given by
\begin{equation}
  G_{c1}(\sqrt{s})=
  \int\!\frac{d^{3}\bm{q}}{(2\pi)^{3}}
           \frac{M_{N}}{E_{N}(\bm{q})}
           \frac{1}{2\omega_{c}(\bm{q})}
           \frac{ [F_{c1}(\bm{q})]^2  } 
%         \frac{F^\prime_{c1}(\bm{q}) \, F_{c1}^\prime(\bm{q}) } %\theta (q^\prime_{\max}-|\bm{q}|)}
	   {\sqrt{s} - E_N(\bm{q}) - \omega_c(\bm{q}) + i \epsilon},
\label{eq:Gc1}
\end{equation}
\begin{equation}
  G_{c2}(\sqrt{s})=
  \int\!\frac{d^{3}\bm{q}}{(2\pi)^{3}}
           \frac{M_{N}}{E_{N}(\bm{q})}
           \frac{1}{2\omega_{c}(\bm{q})}
           \frac{ [F_{c2}(\bm{q})]^2  } 
%           \frac{F^\prime_{c2}(\bm{q}) \, F_{c2}^\prime(\bm{q}) } %\theta (q^\prime_{\max}-|\bm{q}|)}
	   {\sqrt{s} - E_N(\bm{q}) - \omega_c(\bm{q}) + i \epsilon},
\label{eq:Gc2}
\end{equation}
which are regularized with the form factors $F_{c1}(\bm{q})$ and $F_{c2}(\bm{q})$.

We have to define $F_{c1}(\bm{q})$ and $F_{c2}(\bm{q})$.
In Fig.~\ref{fig:2}, let us take the first term corresponding to $\tilde{T}_{11} G_{c1} \tilde{T}_{11}$. In the lower box, we have a nucleon as an external line and the other one appears inside a loop.
One has to see the form factor attached to the diagram of Fig.~\ref{fig:1}(a). 
This is evaluated in Ref.~\cite{Roca:2010tf} when the two particles of the cluster have the same mass, providing a form factor $F_c((\bm{k} - \bm{k}^\prime)/2)$. However, when the masses of the particles in the cluster are different, this form factor is different, and one obtains~\cite{Yamagata-Sekihara:2010kpd} 
\begin{align}
F_{c1} &\rightarrow F_c \left( \frac{m_2}{m_1 + m_2} (\bm{k} - \bm{q}) \right), 
\label{eq:Fc1_0}\\
F_{c2} &\rightarrow F_c \left( \frac{m_1}{m_1 + m_2} (\bm{k} - \bm{q}) \right).
\label{eq:Fc2_0}
\end{align}
Assuming that $\bm{k}$ is small compared to $\bm{q}$ in the loop, we obtain:
\begin{align}
F_{c1}(\bm{q}) & = F_c \left( \frac{m_2}{m_1 + m_2} \bm{q} \right), 
\label{eq:Fc1}\\
F_{c2}(\bm{q}) & = F_c \left( \frac{m_1}{m_1 + m_2} \bm{q} \right).
\label{eq:Fc2}
\end{align}
By using Eq.~\eqref{eq:form_factor0} we can check that, 
for an average value of $|\bm{k}| \sim 150~$MeV/$c$ for the $\bar{K}$ momentum in the cluster distributed isotropically, the changes from Eq.~\eqref{eq:Fc2_0} to Eq.~\eqref{eq:Fc2} 
are about 3\% as an average and much smaller from Eq.~\eqref{eq:Fc1_0} to Eq.~\eqref{eq:Fc1}.
It should be noted that when the loop is closed with the upper box in the diagram, we get an extra,
%\begin{equation}
$F_c\left( \frac{m_2}{m_1 + m_2} (\bm{q} - \bm{k}^\prime) \right) $ or
$F_c\left( \frac{m_1}{m_1 + m_2} (\bm{q} - \bm{k}^\prime) \right) $,
%\end{equation}
with the external $\bm{k}^\prime$ neglected again because it is small. Hence, we get the square of $F_{c1}(\bm{q})$,  $F_{c2}(\bm{q})$ in the loop functions of Eqs.~\eqref{eq:Gc1}, \eqref{eq:Gc2}.
We should also note that in the $t_1 t_2 \tilde{G}$ term, the form factor that appears when the masses of $m_1$ and $m_2$ are different is
\begin{equation}
F_c\left( \bm{q} - \frac{m_2}{m_1 + m_2}\bm{k} - \frac{m_1}{m_1 + m_2}\bm{k}^\prime \right)
\;\rightarrow\; F_c(\bm{q}), 
\end{equation}
neglecting the momenta $\bm{k}$ and $\bm{k}^\prime$ of the external particles.
To justify the use of the same form factors in the iterated diagrams, one can see details in Appendix~A of Ref.~\cite{Malabarba:2024hlv}.

The sum of the terms in Eq.~\eqref{eq:T_sum} can be done in a compact way by writing the matrix,
\begin{equation}
  \tilde{T}_{ij} \;=\;
  \begin{pmatrix}
    \tilde{T}_{11} & \tilde{T}_{12}\\
    \tilde{T}_{21} & \tilde{T}_{22}
  \end{pmatrix},
\end{equation}
and adding the $\tilde{T}_{ij}$ terms obtained prior to the coherent propagation, we have 
\begin{equation}
  \tilde{T}^{\prime (2)}_{ij}
  = \tilde{T}_{ij} + \tilde{T}_{i \ell} \,G_{c}\,\tilde{T}_{\ell j},
\label{eq:Ttilde0}
\end{equation}
where
\begin{equation}
G_c \equiv 
\begin{pmatrix}
G_{c1} & 0 \\
0 & G_{c2} 
\end{pmatrix}.
\end{equation}
Eq.~\eqref{eq:Ttilde0} is the expansion to second order of the Lippmann–Schwinger equation.

The full unitarization (Lippmann–Schwinger equation) means that the kernels $\tilde{T}_{ij}$ are iterated through multiple $n\bar{D}_{s0}^{*}$ intermediate steps, instead of the one step that one has in Eq.~\eqref{eq:Ttilde0}.  
Similar to the Lippmann–Schwinger equation, this is accomplished via
\begin{equation}
  \tilde{T}^{\prime}_{ij}
  = \tilde{T}_{ij} + \tilde{T}_{ij} \,G_{c}\,\tilde{T}^\prime_{ij}.
\label{eq:Ttilde}
\end{equation}
which in the matrix form  has the solution,
\begin{equation}
  \tilde{T}^{\prime}
=  [ 1-\tilde{T}\,G_{c} ]^{-1}\,\tilde{T}.
\label{eq:Ttilde1}
\end{equation}
Finally, our unitary matrix has the form
\begin{align}
  \tilde{T}^\prime_{\text{sum}}
  &= \tilde{T}^\prime_{11}+\tilde{T}^\prime_{12}+\tilde{T}^\prime_{21}+\tilde{T}^\prime_{22} \nonumber\\
  &= 
     \frac{ \tilde{T}_{11}+2\tilde{T}_{12}+\tilde{T}_{22}
        +\bigl(\tilde{T}^{2}_{12}-\tilde{T}_{11}\tilde{T}_{22}\bigr) (G_{c1}+G_{c2}) }
        { 1-\tilde{T}_{11}G_{c1}-\tilde{T}_{22}G_{c2}
        -\bigl(\tilde{T}^2_{12}-\tilde{T}_{11}\tilde{T}_{22}\bigr)G_{c1} G_{c2}}.
\label{eq:Ttilde_prime}
\end{align}

In order to test the elastic unitarity, we look at the linear terms in $i\,q_{\text{cm}}$ that emerge from $\tilde{T}_{\text{sum}}^\prime$ in Eq.~\eqref{eq:Ttilde_prime} close to threshold.
Since, close to threshold,
\begin{equation}
  \sqrt{s}
    = M_{n}+M_{c}+\frac{q_{\text{cm}}^{2}}{2\mu},
  \quad
  \mu = \frac{M_{n} M_{c}}{M_{n}+M_{c}},
\end{equation}
with $\mu$ the $n\bar{D}_{s0}^*$ reduced mass,
we check the exact unitarity by looking if $F_{\text{uni}}=1$, where
\begin{align}
F_{\text{uni}} = \lim_{q_{\text{cm}} \to 0}
\Biggl[
  &\frac{8\pi \left( \sqrt{s}_{\text{th}} + \frac{q_{\text{cm}}^2}{2\mu} \right) }{2M_N}
    \left(\tilde{T}^\prime_{\text{sum}} \left( \sqrt{s}_{\text{th}} + \frac{q_{\text{cm}}^2}{2\mu} \right)
\right)^{-1}
\nonumber \\
  &\quad -
  \frac{8\pi \sqrt{s_{\text{th}}}}{2M_N}
    \left( \tilde{T}'_{\text{sum}} (\sqrt{s}_{\text{th}}) \right)^{-1} 
\Biggr]
\big/ (i q_{\text{cm}}),
\label{eq:Funi}
\end{align}
where the subscript ``th’’ refers to the $n \bar{D}_{s0}^*(2317)$ threshold ($M_N \equiv M_n$).
Note that
\begin{align}
 G_{ci}(\sqrt{s}) & =\text{Re}G_{ci}(\sqrt{s})+i\,\text{Im}G_{ci}(\sqrt{s}),\\
\tilde{G}(\sqrt{s}) &=\text{Re}\tilde{G}(\sqrt{s})+i\,\text{Im}\tilde{G}(\sqrt{s}).
\end{align}
and we have
\begin{align}
\text{Im}G_{ci}
=-\frac{2M_N}{8\pi\sqrt{s}}\;q_{\text{cm}}\;F_{ci}^2(q_{\text{cm}}), \\
\text{Im}\tilde{G}
=-\frac{2M_N}{8\pi\sqrt{s}}\;q_{\text{cm}}\;F_c(q_{\text{cm}}),
\end{align}
with
\begin{equation}
  F_c(q_{\text{cm}})\;\simeq\;
  1-\frac{1}{6}\,\langle r^{2}\rangle\,q_{\text{cm}}^{2},
\end{equation}
with $\langle r^{2}\rangle$ the mean–square radius of the cluster wave function and similar expansions for $F_{ci}(q_{\text{cm}})$.
Hence, up to linear terms in $q_{\text{cm}}$, $\text{Im}\, G_{ci}$ ($i=1,2$) and $\text{Im}\, \tilde{G}$ are identical.
Then, close to threshold, an expansion in terms of $q_{\text{cm}}$ goes as
\begin{equation}
   \tilde{T}^\prime_{\text{sum}} (q_{\text{cm}} )
  \simeq
   \tilde{T}^\prime_{\text{sum}}|_{\text{th}}  +i\,\alpha q_{\text{cm}},
\end{equation}
with $\alpha$ a constant.
Hence, Eq.~\eqref{eq:Funi} is well defined. We have checked that $F_{\text{uni}} =1$ within numerical accuracy.
Then the scattering length and effective range are obtained as in Ref.~\cite{Ikeno:2023ojl}.
 \begin{align}
&  a = \frac{2M_N}{8\pi\sqrt{s}}\;
  \tilde{T}^\prime_{\text{sum}} \bigl|_{\text{th}}, \label{eq:a0}\\
&  r_{0} =  \frac{1}{\mu}
  \left[ \frac{\partial}{\partial\sqrt{s}}
    \left(
      -\frac{8\pi\sqrt{s}}{2M_N}\,
      \left(\tilde{T}^\prime_{\text{sum}} \right)^{-1} + i\,q_{\text{cm}}
    \right)
  \right]_{\text{th}}.  \label{eq:r0}
 \end{align}

\subsection{Correlation function}
Following the formalism of Ref.~\cite{Vidana:2023olz}, which is in line with the Koonin–Pratt formalism~\cite{Koonin:1977fh} adapted to our type of amplitudes, the correlation function is written as
\begin{equation}
\begin{aligned}
   C_{n\bar{D}_{s0}^*}(p)
 & = 1  + 4\pi
      \int_{0}^{\infty} dr r^{2}
        S_{12}(r) \cdot \\ 
& \bigg\{  \Big| 
j_{0}(pr) + 
\left[ ( \tilde{T}_{11}^\prime + \tilde{T}_{21}^\prime )\tilde{G}_1(s,r) 
+ (\tilde{T}_{12}^\prime + \tilde{T}_{22}^\prime )\tilde{G}_2(s,r) 
\right] 
\Big|^2 \\
&  -  j_{0}^{2}(pr)   \bigg\},
\label{eq:CF}
\end{aligned}
\end{equation}
where
$\tilde{G}_1 (s,r)$ and $\tilde{G}_2 (s,r)$ are given now by
% \begin{equation}
%   \tilde{G}_{1}(s,r)=
%   \int\!\frac{d^{3}\bm{q}}{(2\pi)^{3}}
%            \frac{1}{2\omega_{\bar{D}_{s0}}(\bm{q})}
%            \frac{M_{N}}{E_{N}(\bm{q})}
%            \frac{ j_0(qr) \, F_{c1}( \bm{q})  } 
% 	   {\sqrt{s}  - \omega_{\bar{D}_{s0}}(\bm{q}) - E_N(\bm{q}) + i \epsilon},
% \label{eq:Gtil_CF}
% \end{equation} 
% \begin{equation}
%   \tilde{G}_{2}(s,r)=
%   \int\!\frac{d^{3}\bm{q}}{(2\pi)^{3}}
%            \frac{1}{2\omega_{\bar{D}_{s0}}(\bm{q})}
%            \frac{M_{N}}{E_{N}(\bm{q})}
%            \frac{ j_0(qr) \, F_{c2}( \bm{q})  } 
% 	   {\sqrt{s}  - \omega_{\bar{D}_{s0}}(\bm{q}) - E_N(\bm{q}) + i \epsilon},
% \label{eq:Gtil_CF2}
% \end{equation}
\begin{align}
  \tilde{G}_{1}(s,r)=
  \int\!\frac{d^{3}\bm{q}}{(2\pi)^{3}} &
           \frac{1}{2\omega_{\bar{D}_{s0}}(\bm{q})}
           \frac{M_{N}}{E_{N}(\bm{q})} \cdot \nonumber\\
           & \frac{ j_0(qr) \, F_{c1}( \bm{q})  } 
	   {\sqrt{s}  - \omega_{\bar{D}_{s0}}(\bm{q}) - E_N(\bm{q}) + i \epsilon},
\label{eq:Gtil_CF}
\end{align}
%\end{equation}
\begin{align}
  \tilde{G}_{2}(s,r)=
  \int\!\frac{d^{3}\bm{q}}{(2\pi)^{3}} &
           \frac{1}{2\omega_{\bar{D}_{s0}}(\bm{q})}
           \frac{M_{N}}{E_{N}(\bm{q})} \cdot \nonumber\\
           & \frac{ j_0(qr) \, F_{c2}( \bm{q})  } 
	   {\sqrt{s}  - \omega_{\bar{D}_{s0}}(\bm{q}) - E_N(\bm{q}) + i \epsilon},
\label{eq:Gtil_CF2}
\end{align}
with $F_{c1} (\bm{q})$ in Eq.~\eqref{eq:Fc1} and and $F_{c2} (\bm{q})$ in Eq.~\eqref{eq:Fc2}.
The reason for the new form of Eq.~\eqref{eq:CF} is that the scattering matrix needed here to reproduce the wave function is the half-off-shell $\tilde{T}^\prime$ matrix. One line corresponds to momentum $\bm{p}$ and the other to the variable $\bm{q}$ in the loop.
In our case, this would be the on-shell $\tilde{T}^\prime$ matrix multiplied by the form factors $F_{c1}(\bm{p}) F_{c1}(\bm{q})$ (or $F_{c2}(\bm{p}) F_{c1}(\bm{q})$), $F_{c2}(\bm{p}) F_{c2}(\bm{q})$ (or $F_{c1}(\bm{p}) F_{c2}(\bm{q})$)
but we assume $F_{c1}(\bm{p}) = F_{c2}(\bm{p}) \simeq 1$, 
because $\bm{p}$ is small, and hence inside the loop function we have $F_{c1}(\bm{q})$ or $F_{c2}(\bm{q})$, depending on whether the $\tilde{T}^\prime$ matrix finishes with the interaction in particle-1 ($\tilde{T}_{11}^\prime + \tilde{T}_{21}^\prime $)
or in particle-2 $(\tilde{T}_{12}^\prime + \tilde{T}_{22}^\prime )$.

The momentum $p$ of the particles in the rest frame of the pair is
\begin{equation}
  p =  \frac{\lambda^{1/2} (s,M_{N}^{2},M_{\bar{D}_{s0}^*}^{2})}
       {2\sqrt{s}}, 
\end{equation}
with $\lambda$ the K\"{a}llen function. $S_{12}(r)$ in Eq.~\eqref{eq:CF} is the source function which encodes the probability that the two particles of the correlation function are produced at relative distance~$r$. It is usually parametrized as
\begin{equation}
 S_{12}(r) =  \frac{1}{\bigl(4\pi R^{2}\bigr)^{3/2}}\,
  \exp \left( -\frac{r^{2}}{4R^{2}} \right), 
\end{equation}
and the value of $R$ is tied to the type of experiment used to measure the correlation function,
$R\simeq 1$~fm for $pp$ ultrarelativistic collisions, $R\simeq 5$~fm for heavy-ion collisions.
We shall thus take three values of $R$ around $R=1$~fm.

A recent paper~\cite{Epelbaum:2025aan} discusses that the source function is not universal and depends on the model used to describe the two-particle scattering matrix because of different off-shell extrapolations of the
various models employed.
While this statement is certainly correct, the use of realistic models
shows that the uncertainties are at the level of $2$–$5\%$ in the case
of $NN$ correlation functions~\cite{Gobel:2025afq}, and $2$--$3\%$ when studying meson–baryon correlation functions~\cite{Molina:2025lzw}, so we do not worry about this issue here.
We should note that the factor $j_{0}(qr)$ in Eqs.~\eqref{eq:Gtil_CF},\eqref{eq:Gtil_CF2}  already acts as a regulator of $\tilde{G}_i$ ($i=1,2$) and the use of the extra  $F_{ci}(\bm{q})$ in Eqs.~\eqref{eq:Gtil_CF}, \eqref{eq:Gtil_CF2} has a minor extra effect.
As shown in Ref.~\cite{Molina:2025lzw}, 
where instead of the form factors $F_{ci}(\bm{q})$ one has a step function $\theta(q_{\text{max}} - |\bm{q}|)$,
a change of $q_{\max}$ from $630$~MeV to $1000$~MeV produces changes in the correlation function of the order of $2\%$.

\section{Numerical Results}\label{sec:result}
We show here results for the amplitude obtained below threshold, the threshold scattering parameters, $a$ and $r_0$, and the correlation functions.

\subsection{Resonance state below the $n \bar D_{s0}^*(2317)$  threshold}
\begin{figure}[!tb]
\centering
\includegraphics[width=0.5\textwidth]{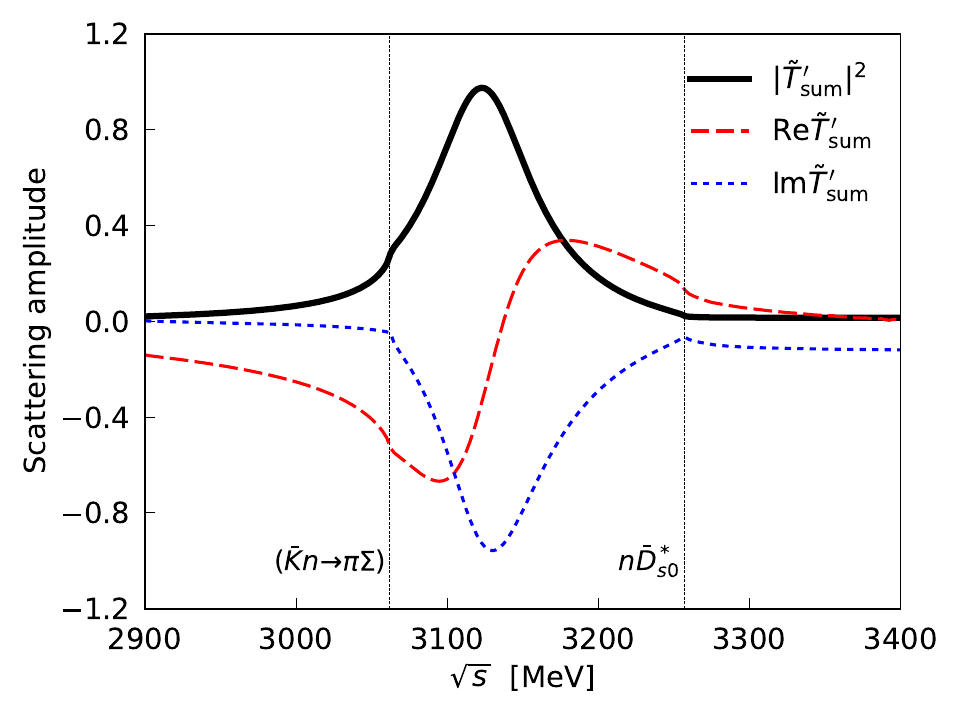}
\caption{ The scattering amplitude $\tilde{T}^\prime_{\text{sum}}$ for the $n \bar D_{s0}^*(2317)$ interaction as a function of $\sqrt{s}$.
The red dashed line shows the real part of the amplitude $\tilde{T}^\prime_\text{sum}$, the blue dotted line the imaginary part of $\tilde{T}^\prime_\text{sum}$, and the black solid line the modulus squared of $\tilde{T}^\prime_\text{sum}$. 
The vertical lines indicate the $n \bar D_{s0}^*(2317)$  threshold and the appearance of the $\pi \Sigma$ channel in $\bar K n \to \pi \Sigma$, corresponding to $s_2$ of Eq.~\eqref{eq:s1s2}. 
}
\label{fig:Tsum}
\end{figure}

We take the $\tilde{T}^\prime_{\text{sum}}$  amplitude and look at it below the $n \bar D_{s0}^*(2317)$ threshold. We show the results in Fig.~\ref{fig:Tsum}. The real and imaginary parts of the amplitude have the shape of a typical Breit Wigner amplitude, with the imaginary part having a peak and the real part going from negative to positive at the peak of the imaginary part.  The peak of $|\tilde{T}^\prime_{\text{sum}}|^2$ appears around 130 MeV below threshold and has a width of about 80 MeV. 
This is already interesting because if the $D_{s0}^*(2317)$ state were an ordinary state, and there was a $n \bar{D}_{s0}^*(2317)$ bound state, we should expect it to be very narrow, with a width coming from the natural decay of the $\bar D_{s0}^*(2317)$ into $\bar D_s \pi$, which is very small because the decay is isospin forbidden. Instead we obtain a width of the order of 80 MeV, far larger than the decay width of the $\bar D_{s0}^*(2317)$, which is smaller than 3.8 MeV, according to the PDG~\cite{ParticleDataGroup:2024cfk} (of the order of 100 KeV in theoretical calculations~\cite{Guo:2018kno}).

  In the case of the $D_{s0}^*(2317)$ as a molecular state of $DK$, the interaction of the neutron with ($\bar D \bar K$) involves the interaction of the neutron with $\bar K$ which gives rise to the two $\Lambda(1405)$ states~\cite{Oller:2000fj,Jido:2003cb}.
The width originates from the decay widths of these resonances, which comes from decay to $\pi \Sigma$. 
The $\pi \Sigma$ channel according to the relationship of $s_2$ and $s$ in Eq.~\eqref{eq:s1s2}, opens up at $\sqrt{s} = 3063$~MeV, and is visible in Fig.~\ref{fig:Tsum} as a kink in the real  and imaginary parts of $\tilde{T}^{\,\prime}_{\text{sum}}$.  
The thresholds for $n\bar K$ and $n\bar{D}$ appear for $\sqrt{s}$ above the $n \bar{D}^*_{s0}(2317)$ threshold and lead to no visible structure.
We also find that the binding is larger than what we would expect from the binding of the two $\Lambda(1405)$ (around 45 MeV for the lower mass $\Lambda(1405)$ state). An explanation on why one obtains bigger binding in the three body system than in the two body one is provided in the appendix of Ref.~\cite{Montesinos:2024eoy} 
and has to do with the relationship of the invariant mass of the external particle and the cluster, and the invariant mass of the external particle with one particle of the cluster. Needless to say that the observation of this peak in some reaction would tell us much about the nature of the $D_{s0}^*(2317)$ state.

Since the width of the state that we find is related to the $\bar{K} N \to \pi \Sigma$ transition, one should look at $\bar{D} \pi \Sigma$, looking for a peak in the mass distribution of these three particles. This is a standard technique in the LHCb collaboration using $pp$ collisions at high energy and selecting trios of particles to identify their origin. 
This is the way heavy hadrons decaying to three particles are identified~\cite{LHCb:2015yax,LHCb:2020bls}.

\subsection{Scattering length and effective range for $n \bar D_{s0}^*(2317)$ scattering}
By looking at the amplitude $\tilde{T}^\prime_{\text{sum}}$ close to threshold, we can determine  the scattering length and effective range for $n \bar D_{s0}^*(2317)$ scattering using Eqs.~\eqref{eq:a0}, \eqref{eq:r0}.
The obtained scattering length $a$ and effective range $r_0$ are 
\begin{align}
 a &= 0.60 - i\, 0.29~\text{fm}, \\
 r_0 &= 1.14 - i\, 0.11~\text{fm}.
\end{align}
These numbers are similar as those obtained in Ref.~\cite{Encarnacion:2025lyf}, but the imaginary part of $r_0$ has opposite sign to the one obtained in Ref.~\cite{Encarnacion:2025lyf} for  $p f_1(1285)$ scattering.

\subsection{Correlation functions}
\begin{figure}[!tb]
\centering
\includegraphics[width=0.5\textwidth]{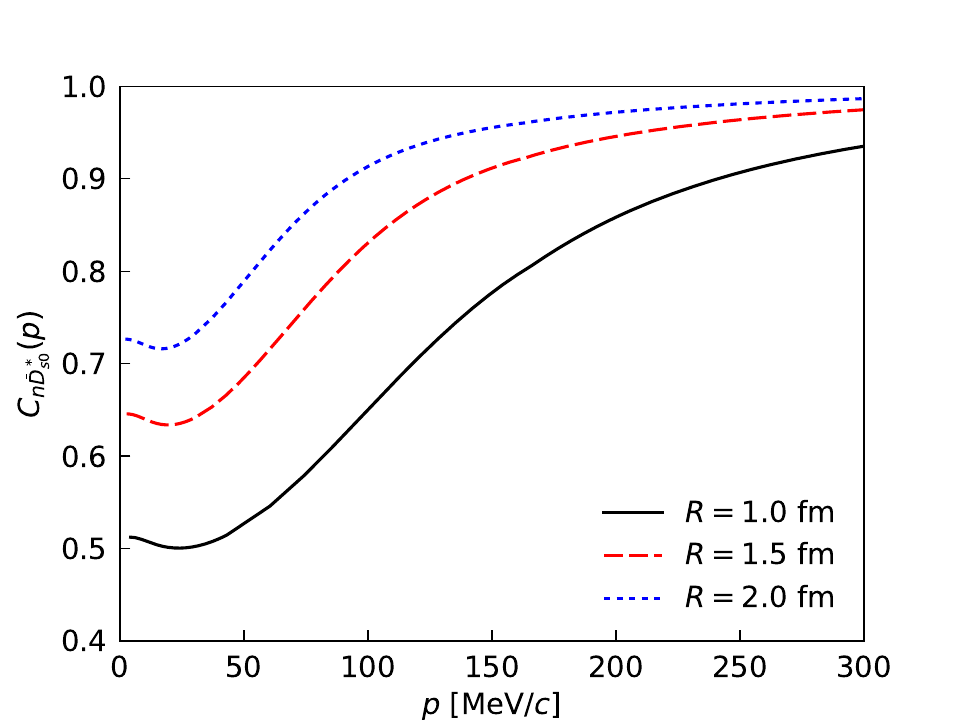}
\caption{The correlation function of $n \bar{D}_{s0}^*(2317)$ for the different value of $R$. }
\label{fig:CF}
\end{figure}

Next we look at the $n \bar{D}_{s0}^*(2317)$ correlation function of Eq.~\eqref{eq:CF}. We plot the results in Fig.~\ref{fig:CF} for three values of the $R$ parameter of the source function. By taking $R=1$~fm for discussion, we see that we obtain a typical correlation function for the case when one has some bound state below threshold. 
One can see that the shape obtained in Fig.~\ref{fig:CF} is similar to Fig.~3(d3) of Ref.~\cite{Liu:2023uly}, corresponding to the case labeled there as 
strongly bound scenario with real potentials (see also Ref.~\cite{ExHIC:2017smd}), or the $K^+ \Sigma^0$ correlation function of Ref.~\cite{Molina:2023jov}, which is tied to the $N^*(1535)$ state  with a width more in line with the state found here.
The shape and strength are very similar to those obtained in Ref.~\cite{Encarnacion:2025lyf}  for the correlation function of  $p f_1(1285)$ scattering, where also a resonant state below threshold was found. Experimentally the correlation function would be measured by selecting a neutron and then the pair $\bar D_s \pi$ with invariant mass around the  2317 MeV of the $\bar D_{s0}^*(2317)$. 

\section {Conclusions}\label{sec:discussion}
We have studied the interaction of a neutron with the $\bar D_{s0}^*(2317)$ resonance and evaluated the scattering parameters close to threshold, $a$, $r_0$, plus the correlation function for small neutron momenta in the $n \bar D_{s0}^*(2317)$ rest frame. We have also looked at the scattering amplitude below threshold and have found a clean resonant state with binding about 130 MeV and a width of about 80 MeV. This is a distinctive feature due to the nature assumed for the $\bar D_{s0}^*(2317)$ state as mostly a ($\bar{D} \bar{K}$) bound state. 

    For the theoretical frame, we started using the Fixed Center Approximation to the Faddeev equations, where the $\bar D_{s0}^*(2317)$ was assumed to be the cluster and the neutron the external particle. Yet, it was shown that the standard amplitude of the FCA does not satisfy elastic unitarity close to threshold, which is essential to determine the effective range and the correlation function. The needed corrections to the standard formula were obtained by looking at the analogous case of particle-nucleus interaction, where one obtains first an optical potential, by looking at the dynamics of the problem, and then one solves the Schr\"{o}dinger equation (Lippmann Schwinger equation for scattering). We identified the terms that had to be iterated through a coherent propagation of the neutron and the cluster, and finally we obtained a formula that contains the basic features of the dynamics and fulfills exactly unitarity at threshold. This is a novel aspect of the work, with a necessary improvement of the FCA for application to scattering above threshold, which opens the door to study many similar problems of scattering of external particles with molecular states of any type. 
    
    Coming back to the state found below threshold, since the width comes from decay of $\bar K n$ to $\pi \Sigma$, the state should be investigated in the mass distribution of $\bar D \pi \Sigma$. 
This is a distinctive feature of the molecular picture because in the case in which the $\bar D_{s0}^*(2317)$  was an elementary particle, if there would be a $n \bar D_{s0}^*(2317)$ bound state, it should be found in the $\bar D_s \pi n$ system and one should expect a much smaller width than in the case that the $\bar D_{s0}^*(2317)$ is a molecular state.

    The measurement of the correlation function, the scattering length and the effective range, together with the resonant state below threshold, would provide very useful information to test the predictions based on the molecular picture of the $D_{s0}^*(2317)$ resonance, eventually giving it extra support. We are looking forward to such experiments in the future, which undoubtedly will improve our understanding of the hadronic spectrum.

\section*{Acknowledgments}
The work was partly supported by JSPS KAKENHI Grant Number JP24K07020.
This work is partly supported by the Spanish Ministerio de Economia y Competitividad (MINECO) and European FEDER funds under Contracts No. FIS2017-84038-C2-1-P B, PID2020- 112777GB-I00, and by Generalitat Valenciana under contracts PROMETEO/2020/023 and CIPROM/2023/59.This project has received funding from the European Union Horizon 2020 research and innovation programme under the program H2020- INFRAIA2018-1, grant agreement No. 824093 of the STRONG-2020 project. This work is supported by the Spanish Ministerio de Ciencia e Innovacion (MICINN) under contracts PID2020-112777GB-I00, PID2023-147458NBC21 and CEX2023-001292-S.

\appendix
\section{ $\bar{K}N$ and $\bar{D}N$ scattering amplitudes}\label{sect:app}
We need the following amplitudes.

%\subsection*{\texorpdfstring{$\bar{K}N$}{Kbar N}, $I=0$}
\subsection{$\bar{K}N$, $I=0$}

We use the chiral unitary approach of Ref.~\cite{Oset:1997it} using the coupled channels, $\bar{K}N$, $\pi\Sigma$, $\eta\Lambda$, and $K\Xi$. The transition potentials are given by
\begin{equation}
  V_{ij} = -\frac{1}{4f^{2}} D_{ij} (k^{0}+k^{\prime 0}),
  \quad  f = 1.15 \times 93~\text{MeV},
\label{eq:V_KNI0}
\end{equation}
with $k^{0}$ and $k^{\prime 0}$ the energies of the mesons,
\begin{equation}
  k^{0}
  = \frac{s+m_{i}^{2}-M_{i}^{2}}{2\sqrt{s}},
  \quad
  k^{\prime 0}
  = \frac{s+m_{j}^{2}-M_{j}^{2}}{2\sqrt{s}},
\end{equation}
with $m_{i},M_{i}$ the initial meson and baryon masses for each transition and $m_{j},M_{j}$ the same ones for the final meson–baryon state.  
The coefficients $D_{ij}$ are given in Table 2 of Ref.~\cite{Oset:1997it}.  
The scattering matrix is obtained in matrix form as  
\begin{equation}
  T = \bigl[1-V G\bigr]^{-1}\,V . 
\label{eq:TVG}
\end{equation}

%----------------------------------------------------------
\subsection{$\bar{K}N$, $I=1$}
Now the coupled channels are  
$\bar{K}N$, $\pi\Sigma$, $\pi\Lambda$, $\eta\Sigma$, and $K\Xi$ and the transition potentials are given by
\begin{equation}
  V_{ij}  = -\frac{1}{4f^{2}}\,
      F_{ij}\, (k^{0}+k^{\prime 0}) ,
  \quad  f = 1.15 \times 93~\text{MeV},
\label{eq:V_KNI1}
\end{equation}
with the coefficients $F_{ij}$ given in Table 3 of Ref.~\cite{Oset:1997it}.

The amplitudes $V_{ij}$ of Eq.~\eqref{eq:V_KNI0} are based on the lowest order chiral Lagrangian (see Ref.~\cite{Ecker:1994gg}) and the framework to evaluate $T_{ij}$ requires just one parameter to regularize the $G$ function of Eq.~\eqref{eq:TVG}. Further developments incorporate higher--order Lagrangians involving many parameters that are fitted to experimental data \cite{Ikeda:2012au,Guo:2012vv,Mizutani:2012gy,Mai:2014xna,Feijoo:2015yja,Feijoo:2018den}, and one aims at describing data at higher energies. Yet, as can be seen in Ref.~\cite{Oset:1997it}, the description of $\bar{K}N$ data is exceptionally good already at the lowest order level of chiral Lagrangians used here, in the range of energies where we are concerned in the present work.

%----------------------------------------------------------
\subsection{$\bar{D}N$, $I=0,1$ }
The $\bar{D}N$ interaction is equivalent to the $KN$ one, which is also studied in Ref.~\cite{Oset:1997it}.
Given the equivalent of $(K^+, K^0) \to (\bar{D}^0, D^-)$, one finds then
\begin{equation}
  V_{ij}  = -\frac{1}{4f^{2}}\,
      L_{ij}\, (k^{0}+k^{\prime 0}) ,
  \quad  f = 93~\text{MeV},
\end{equation}
with the coefficients $L_{ij}$ given in Table~\ref{tab:Lij}, which immediately leads to
\begin{align}
  V^{(I=0)} = -\frac{1}{4f^{2}} L^{(0)}\,
      (k^{0}+k^{\prime 0}), \\
  V^{(I=1)} = -\frac{1}{4f^{2}}\,
      L^{(1)}\,
      (k^{0}+k^{\prime 0}),
\end{align}
with $L^{(0)} = 0$ and $L^{(1)} = -2$.
We see that for $I=0$ the interaction is zero, whereas for $I=1$ it is repulsive.

\begin{table}[!ht]
  \centering
  \caption{Coefficients $L_{ij}$ in the $\bar{D}^{0}n$–$D^{-}p$ basis}
  \label{tab:Lij}
  \begin{tabular}{c|cc}
    \hline
   ~~~ $L_{ij}$    ~~~     &  ~~ $\bar{D}^{0} n$ ~~&  ~~$D^{-} p$ ~~ \\ \hline
   ~~~ $\bar{D}^{0} n$  ~~~&  $-1$  & $-1$ \\
   ~~~ $D^{-} p$    ~~~ & $-1$ & $-1$ \\ \hline
  \end{tabular}
\end{table}

\bibliography{ref_nDs0bar}

\end{document}